# Free-floating planets: a viable option for panspermia

Hector Javier Durand-Manterola

Departamento de Ciencias Espaciales, Instituto de Geofísica
Universidad Nacional Autónoma de México
E-mail: hdurand@geofisica.unam.mx

**Abstract**

Context: Genomic complexity can be used as a clock with which the moment in which life originated can be measured. Some authors who have studied this problem have come to the conclusion that it is not possible that terrestrial life originated here and that, in reality, life originated giga-years ago, before the solar system existed. If we accept this conclusion there is no other option than to admit that panspermia is something viable. However, much has been studied about the possibility that simple life forms (SLF) go from one system to another and the conclusion has been that transit times are so long and the environment of the interstellar space is so hostile that it makes this transit highly improbable.

Objective: The goal of this study is to propose a viable hypothesis for the transport of SLF from one planetary system to another.

Method: During the formation period of a planetary system giant planets can eject planets the size of the Earth, or larger, turning them into free-floating planets in interstellar space. These free-floating planets have also been called free floaters. If a free floater, which has developed life, enters a lifeless planetary system, it can seed the worlds of this system with SLF dragged by the stellar wind from one planet to another or by great impacts on the free planet. To support this hypothesis, I calculate the probability that one free floater reaches the planets zone of a planetary system, and also it was calculated the time it remains within the planetary zone in order to see if there is enough time to seed the host system.

Results: The probability of a free floater in the galaxy, within the region of the Sun, entering the planet zone of a system is $2.8 \times 10^{-4}$, i.e., that ~3 of 10,000 free planets manage to enter some planetary system. At the galactocentric distance from the Sun I calculated that there are 21,495 free floaters floating around the galactic center. Hence, 6 free-floating planets manage to enter in planetary systems every galaxy rotation. Since the galaxy has rotated 54 times since its formation, then, ~ 324 free floaters have entered some planetary system at the galactocentric distance of the Sun. The period of time they remain in these planetary zones is calculated as between 0 and 138 years. For an interstellar planet which enters a planetary system with an impact parameter of ≤ 10 AU it would take ~ 40 years to travel across the planetary zone.

# 1 Introduction

It has been proposed: that as the live organisms have evolved their genome has become more complex; and it could be interpreted, this genomic complexity, as a measure for the degree of evolution of determined species. The genomic complexity has been defined as the size of the functional and non-redundant genome (Adami et al., 2000; Sharov, 2006). One of these authors, Sharov (2006 y 2010), has studied organisms genomic complexity and uses it as a clock and in this way determined at which moment it was zero, i.e., at which moment life began. He studied several taxonomic groups, measuring their genomic complexity, and performed with these data a lineal regression, which shows that going back in time, genomic complexity decreases. Another result obtained by him is that genomic complexity was still very high at the time of the birth of the Earth (~$1.2 \times 10^5$ bp) (bp = base pairs). Sharov (2006 and 2010) suggests that this result question whether life appeared on our planet or had come from outside in a state of fairly large genomic complexity. Extrapolating the lineal regression back in time, Sharov came to the conclusion that zero for genomic complexity is at 10 ± 3 Ga AP, i.e., 5.4 Ga before the creation of Earth.

The problem with these results is that some authors have studied the possibility that the simple life forms (SLF) passed from one planetary system to another, traveling on meteorites, grains of dust or simply by themselves, stellar wind driven, have concluded that the probability of arriving in a viable state is in the order of zero (Mileikowsky, 2000; Melosh, 2002; Wickramasinghue, 2004). This is because of the fact that the time required for transit is extremely large (in the order of millions of years). During this time SLF receive, from cosmic rays, lethal irradiation. Hence, from this point of view, the panspermia from one planetary system to another seems highly unlikely.

On the other hand, panspermia between bodies of the same planetary system seems more feasible (Wickramasinghe y Wickramasinghe, 2008) mainly because transit times are smaller. Dragged by stellar wind the transit time between planets is of days and in the interior of the meteorites of years.

This work develops a hypothesis which establishes an alternative form in which life can travel from one planetary system to another without any danger of radiation and no matter how long the transit time is. The hypothesis proposed is that the free planets, i.e., those who are not bound to a star, may be the ones on which life appears and later sow the life in the virgin planetary systems. To support this hypothesis, I calculate the probability that a free planet may come to the planetary region of a system, and how long it takes its transit. The results are: the probability, per individual system, of this occurs at the Sun's galactocentric distance is $1.3 \times 10^{-8}$. The probability per rotation of the galaxy is $P = 2.8 \times 10^{-4}$, that is, 3 free-floating planets from every 10,000 pass through a planetary system and could seed it, if they had developed life. I estimate that in the Sun's orbit there are 21,495 interstellar planets so 6 enter every rotation and in the 54 rotations of the galaxy 325 interstellar planets must have entered some planetary system. The time a free planet remains as a guest within a host planetary system, when it enters deep enough, is around 43 years.

## 2 The free planet hypothesis

During the formation of a planetary system the giant planets can eject into interstellar space planets the size of Earth or larger (Lissauer, 1987; Stevenson, 1998; Milone y Wilson, 2008). These interstellar worlds, since they are not bounded to any star and do not receive its heat would have an effective temperature of 30K (Stevenson, 1998). Hence, the escape velocity on the surface would be much greater than the thermal velocity of molecular hydrogen, so, this would not be able to escape and would form atmospheres with pressures of $10^2$ to $10^4$ bars. This would generate opacity, in the $H_2$, induced by pressure; which would allow the surface temperature to exceed the fusion point of water; allowing the existence of liquid water oceans. If the planet had lost its hydrogen before being ejected from its planetary system, then, another possibility would exist. In interstellar space, on a planet of the size of the Earth or larger, the oceans would freeze but would have enough internal heat for geological activity to continue. In the rift zones, where the heat flow is greater, the existing ice would melt and ice caverns would be formed in which there would be covered seas. An example of this possibility is Lake Vostok in Antarctica. This lake is a mass of fresh water 300 km in length and 400 m in depth. It is found at a depth of 3,700 m under a layer of ice in the Antarctic. The Lake formed in a rift valley in which the heat flow kept the water liquid (Carsey and Horvath, 1999). In both cases, heating by $H_2$ or by internal heat flow, in the seas formed life can develop, starting from zero genomic complexity. In the case of oceans frozen on the surface life can appear in the hydrothermal vents or close by, due to the water being full of nutrients from these same vents (Sherwood et al., 1983, pp. 86; Maher y Stevenson, 1988; Pantoja-Alor y Gomez-Caballero, 2004) and the heat emitted by the vents that the SLF may be used for infrared photosynthesis (White et al., 2002). These organisms can evolve and increase their genomic complexity during the journey from the planet in interstellar space without any risk generated by conditions in the space.

At the moment in which the free planet enters a planetary system, becoming a guest planet, the heat from the star makes the hydrogen from the atmosphere evaporate or melts the ice of the guest planet releasing the SLF. These SLF, which may have reached bacterial complexity, can be pulled by the stellar wind and can fall on any of the worlds within the host system. The worlds of this system can be very young like Earth was when life appeared but they would receive a population of highly evolved live beings from the guest planet, maybe with thousands of millions of years of evolution. When a interstellar planet enters a planetary system which is in formation, the amount of debris from the planetary formation will be great; then, the guest planet may be bombarded and would eject into space rocky fragments full of SLF. In the case of Earth, it could be that the seeding of life was done by a guest planet ejected from the first planetary systems formed in the galaxy. The guest planets, entering at escape velocity, would not stay trapped in the orbit of stars in the host system but would follow their route towards interstellar space without disturbing the system again. In the case of free-floating planets heated by hydrogen atmospheres, the entry into a planetary system deeply affects them because the heating due to the star produces the lost of its hydrogen. When the guest planet abandons the host system it faces a cooling which will notoriously affect the living beings developed on it.

**3 Probability that a interstellar planet enters a Planetary System**

To calculate the probability that a free planet passes through a planetary zone, it must be supposed that every star is in the center of a cube with side D, which is the median distance between stars at a distance R from the galactic center. This way, all the galactic space is full of these cubes without any empty space. When a free floater enters one of these cubes it goes through one of the sides of surface $D^2$. If the planet zone around the star in the center of the cube has a transverse section A, then, the probability that the planet enters in this zone is $A/D^2$. If the planetary systems were spherical then their transverse section would be $\pi r^2$, but they are flat and not spherical (all planets are in the plane of the proto planetary disk that formed). Hence, any one system can present different transverse section to two different free-floating planets, depending on the angles at which they approach. For all this, the transverse section will be $\cos(\theta)\pi r^2$, where $\theta$ is the angle that forms the trajectory of the planet with the perpendicular to the plane of the planetary system. Since all the angles are equally likely then, the average value is 45°. Hence, the average transverse section will be $0.7\pi r^2$, where the constant 0.7 is the cosine of 45°.

Thus the probability p of a wandering planet reaches the zone where are the planets of a star, is

$$p = \frac{0.7\pi r^2}{D^2} \qquad (1)$$

Where r is the radius of the planetary zone, D is the median distance between stars and $0.7\pi r^2$ is the average transverse section of the planetary zone.

If the wandering planet follows a circular orbit, with a radius of R, around the center of the galaxy, then the number of planetary systems N which it can find in its trajectory is:

$$N = \frac{2\pi R}{D} \qquad (2)$$

Hence, the total probability P that it crosses a planetary zone will be:

$$P = Np = \begin{cases} \dfrac{1.4\pi^2 r^2 R}{D^3} & \text{if } Np < 1 \\ 1 & \text{if } Np \geq 1 \end{cases} \qquad (3)$$

When the distance to the center of the galaxy increases, the stellar density decreases. Hence, the median distance between stars, D, increases. Thus, the probability, P (equation 3) increases with R but decreases with con $1/D^3$ which is proportional to $1/R^3$, so then, P decreases when R increases as well as $1/R^2$. At the galactocentric distance of our solar system the stellar density is 0.0651 M pc$^{-3}$ (Gilmore y Zeilik, 2000, pp 487). Starting with this value the median distance between stars can be calculated D = 2.5 pc = 8.1 A.L. = 71,058 H. L. (light hours), R = 27,710 AL (IAU 1985 Standard, Trimble, 2000, pp 569) r = 5.906x10$^9$ km (Tholen et al., 2000, pp 294) = 5.47 H.L. With these values p = 1.3x10$^{-8}$ and N = 21,495. Hence, P = 2.8x10$^{-4}$, meaning 3 free planets out of 10,000 pass through a planetary system. Supposing that every planetary system ejects a planet during its formation, then, at the galactocentric distance of the solar system, there must be 21,495 wandering planets. From these, 6 must enter a planetary system every time they orbit the galaxy. At the distance of the Sun the galaxies rotational period is of 2.4x10$^8$ yr (Trimble, 2000, pp 569), hence, in all its history (~13 Ga; Trimble, 2000, pp 572) the galaxy has rotated 54 times. So, 325 free-floating planets will have entered some planetary system.

**4 Transit time**

I call "transit time" to time it takes for the guest planet to travel through the planets zone of a planetary system.

Using r and θ as polar coordinates, with the center on the host star with a mass of M, the movement equations for the guest planet with a mass of m are:

$$m\frac{d^2r}{dt^2} - mr\left(\frac{d\theta}{dt}\right)^2 = -\frac{GMm}{r^2} \quad (5)$$

$$mr\frac{d^2\theta}{dt^2} + 2m\frac{dr}{dt}\frac{d\theta}{dt} = 0 \quad (6)$$

Which with one integration gives the velocity:

$$\frac{dr}{dt} = \sqrt{v_0^2 + \frac{v_0^2 b^2}{r_0^2} - \frac{v_0^2 b^2}{r^2} + \frac{GM}{r} - \frac{GM}{r_0}} \quad (7)$$

This equation can be numerically integrated to give the transit time. Supposing that host stars within the range of 0.8 and 1.4 solar masses, and depending on how much the planet enters, the transit time would be in the range of 0 and 138 years. In a star from the mass of the Sun the transit time is between 0 and 111 years. In figure 1 we can see, if the parameter for the impact is less or equal to 10 UA, passing times vary very little, between 37 and 49 years with an average of 43 years, for any stellar mass. On the other hand transit times are highly dependant on stellar mass, when the parameter of impact is greater than 10 UA.

These times are sufficient to sow the host system planets. In these transit years the guest planet is subjected to stellar wind from the host star and the SLF can be dragged from its atmosphere. On the other hand, if the entry, of the guest planet, occurred during the initial phases of the host system, with a large amount of debris coming from the remains of the protoplanetary disk, then, the guest planet has a high probability of being bombarded and fragments of it, with SLF inside will remain in orbit in the host system. Later these fragments fall into any of the planets of the host system, seeding them.

## 5 Discussion

If the guest planet has a magnetic field the probability of sows, with bacteria or more basic life forms, the planets of the host system decreases. Because the magnetic field prevents the stellar wind interacts with the atmosphere and drag these SLF. However, even in these conditions, the SLF can leave the planet because of the plasma movements within its magnetosphere. Due to the electric fields induced by the stellar wind a plasma flow is produced from the back to the front of the magnetosphere. This flow carries part of the plasmasphere during "peeling" events (Prölss, 2004, pp 251-255). During these events it is possible that the SLF that had reached these altitudes be dragged by the magnetospheric flow and then ejected to interplanetary space where the stellar wind would carry them. In these cases, the dragged material comes from the plasmasphere and not from the mesosphere or the thermosphere. Hence, the difficulty for the SLF to reach the plasmasphere is greater and we can assume that sow will be more difficult. There is a possibility for SLF to climb up to the plasmasphere if the guest planet enters a system that is being formed. Receiving impacts from debris found in its path. These impacts will launch dust with SLF to the upper layers of the atmosphere, to the plasmasphere, and later would be ejected by the peeling events. The magnetic field will not protect the guest planet from the greater impacts, and fragments with SLF which could sow the host system could be ejected; the same would happen with a planet without a magnetic field.

## Conclusions

Although the probability that a wandering planet enters in a planetary system is small (P = $2.8 \times 10^{-4}$) if there are enough of these planets there will be enough encounters so that panspermia becomes an important mechanism in the propagation of life in the universe. The transit times of the guest planets are long enough so that by stellar wind or by asteroid impacts, the SLF may be transferred from the guest planet to the host planetary system.

The hypothesis of the free planets solves the problems presented by panspermia in meteorites, comets or the SLF alone because it do not have the problem of exposure to space.

## Referencias